\documentclass[11pt,fleqn]{article}
\usepackage{amssymb}

\usepackage{amsmath}

\usepackage{amsmath,amssymb,amsthm}

\begin{document}

\begin{center}
{\Large \textbf{On Reduction of Multi-Dimensional Non-Linear \\
Wave  Equation to Two-Dimensional Equations\footnote{Reports
 of the Acad. Sci. of Ukraine. Ser. À, Phys., Math and Tech. Sciences,
1990, No.~8, p.~31--33.}}}

\ \vskip 3pt {\large \textbf{W.I. Fushchych, I.A. Yehorchenko}}
\vskip 6pt {Institute of Mathematics of the National Academy of
Sciences of Ukraine, 3~Tereshchenkivs'ka~Str., 01601 Kyiv-4,
Ukraine.}\\ E-mail: iyegorch@imath.kiev.ua
\end{center}

\begin{abstract}
{A condition of reduction of multidimensional wave equations to
the two-dimensional equation is studied, and the necessary
conditions of compatibility and exact solutions of the resulting
d'Alembert--Hamilton system are obtained.}
\end{abstract}

\medskip

{\bf 1.} We find solutions of the nonlinear wave equation
\begin{equation}
\Box u=F(u),\vspace{1mm}\\
\end{equation}
\[
\Box\equiv \partial^2_{x_0}-\partial^2_{x_1}-\cdots
-\partial^2_{x_n}, \quad u=u(x_0, x_1, \ldots, x_n)
\]
by means of ansatz [1--4]
\begin{equation}
u=\varphi(y,z), \end{equation} where $y$, $z$ are new variables.
The substitution of (2) into (1) leads to equation
\begin{gather}
\varphi_{yy}y_{\mu}y_{\mu}+2\varphi_{yz}z_{\mu}y_{\mu}+
\varphi_{zz}z_{\mu}z_{\mu}+\varphi_y \Box y+\varphi_z \Box
z=F(\varphi)\vspace{1mm}
\end{gather}
\[
\left(y_{\mu}=\frac{\partial y}{\partial x_{\mu}}, \
\varphi_y=\frac{\partial \varphi}{\partial y}\right),
\]
whence we obtain a system of equations:
\begin{gather}
y_\mu y_{\mu}=r(y,z), \quad y_{\mu} z_{\mu}=q(y,z), \quad
z_{\mu}z_{\mu}=s(y,z), \vspace{1mm}\\
\Box y=R(y, z), \quad \Box z=S(y, z) \nonumber.
\end{gather}

The system (4) is a condition of reduction of the
multi-dimensional wave equation (1) to the two-dimensional
equation (3) by means of ansatz (2).

Such reduction is of interest as solutions of two-dimensional
partial differential equations, including non-linear ones, may be
studied to a larger extent than solutions of multi-dimensional
equations.

E.g. let $y_{\mu}y_{\mu}=-z_{\mu}z_{\mu}=1$, $z_{\mu}y_{\mu}=\Box
y= \Box z=0$. Then (3) has the form
\[
\varphi_{yy}-\varphi_{zz}=F(\varphi).
\]

If $F(\varphi)=\sin \varphi$, then the reduced equation has
soliton solutions. If $F(\varphi) = \exp \varphi$, it has a
general solution.

{\bf 2.} Let us formulate the necessary conditions for
compatibility of the d'Alembert--Hamilton system for two
functions.

The system (4), depending on the sign of the expression $rs-q^2$,
may be reduced by local transformations to one of four types:

1) elliptic case: $rs-q^2>0$, $v=v(y,z)$ is a complex-valued
function,
\begin{gather}
\Box v=V(v, v^*), \quad \Box v^*=V^*(v, v^*),\vspace{1mm} \nonumber \\
v^*_{\mu}v_{\mu}=h(v, v^*), \quad v_{\mu}v_{\mu}=0, \quad
v^*_{\mu} v^*_{\mu}=0
\end{gather}
(the reduced equation is of elliptic type);

2) hyperbolic case: $rs-q^2 < 0$, $v=v(y, z)$, $w=w(y, z)$ are
real functions,
\begin{gather}
\Box v=V(v, w), \quad \Box w = W(v, w),\vspace{1mm} \nonumber\\
w_{\mu}w_{\mu}=h(v, w), \quad v_{\mu}v_{\mu}=0, \quad w_{\mu}
w_{\mu}=0
\end{gather}
(the reduced equation is of hyperbolic type);

3)   parabolic case: $rs-q^2=0$, $r^2+s^2+q^2\not=0$, $v(y,z)$,
$w(y,z)$ are real functions,
\begin{gather}
\Box v=V(v,w), \quad \Box w = W(v,w),\vspace{1mm} \nonumber \\
v_{\mu}w_{\mu}=0, \quad v_{\mu}v_{\mu}=\lambda
 \ (\lambda=\pm 1), \quad w_{\mu} w_{\mu}=0
\end{gather}
(if $W\not=0$, the reduced equation is of parabolic type);

4) first-order equation $r=s=q=0$: $y \to v$, $z \to w$
\begin{gather}
 v_{\mu}v_{\mu}=w_{\mu} w_{\mu}=v_{\mu}w_{\mu}=0,\vspace{1mm} \nonumber\\
\Box v=V(v, w), \quad \Box w=W(v, w).
\end{gather}

Compatibility analysis of the d'Alembert--Hamilton system.
\begin{gather}
\Box u=F(u), \quad u_{\mu} u_{\mu}=f(u)
\end{gather}
in three-dimensional space was done by Collins in [5]. Necessary
conditions of compatibility of system (9) for four independent
variables were studied in [6].

Let us formulate necessary conditions for compatibility of the
systems (5)--(8).

\smallskip

\noindent {\bf Theorem 1.} {\it System (5) is compatible only in
the case if
\[
V=\frac{h(v,v^*)\partial_{v^*}\Phi}{\Phi}, \quad
\partial_{v^*}\equiv \frac{\partial}{\partial v^*},
\]
is $\Phi$ an arbitrary function for which the following condition
is fulfilled}
\[
(h\partial_{v^*})^{n+1}\Phi=0.
\]

\noindent {\bf Theorem 2.} {\it System (6) may be compatible only
in the case if
\[
V=\frac{h(v,w)\partial_{w}\Phi}{\Phi}, \quad
W=\frac{h(v,w)\partial_v\Psi}{\Psi},
\]
where the functions $\Phi$, $\Psi$ satisfy the following
conditions}
\[
(h\partial_v)^{n+1}\Psi=0, \quad (h\partial_w)^{n+1}\Phi=0.
\]

\noindent {\bf Theorem 3.} {\it System (7) is compatible only in
the case if}
\[
V=\frac{\lambda \partial_v \Phi}{\Phi}, \quad
\partial_v^{n+1}\Phi=0, \quad W\equiv 0.
\]

\smallskip

System (8) is compatible in the case if $V=W\equiv 0$.

Proof of these theorems is done by means of application of the
lemmas adduced in [6], and of the well-known Hamilton-Cayley
theorem in accordance to which a matrix is root of its
characteristic equation.

\smallskip

\noindent {\bf Note 1.} Equation (5) may be written for a pair of
real functions $\omega={\rm Re}\, v$, $\sigma = {\rm Im}\,v$.
However, in this case the compatibility conditions would look too
cumbersome.

\smallskip

\noindent {\bf Note 2.} Transition from (4) to (5)--(8) is
convenient only with respect to investigation of compatibility.
The sign of the expression $rs-q^2$ may alternate for various $y$,
$z$, and the transition is considered only in the area where this
sign is constant.

\smallskip

{\bf 3.} Let us adduce explicit solutions of systems of type (4)
and reduced equations. Parameters $a_{\mu}$, $b_{\mu}$, $c_{\mu}$,
$d_{\mu}$ $(\mu = \overline{0, 3})$ satisfy the following
conditions:
\[
-a^2=b^2=c^2=d^2=-1\quad (a^2\equiv a^2_0-a^2_1-\cdots - a^2_3),
\vspace{1mm}\\
ab=ac=ad=bc=bd=cd=0;
\]
$y$, $z$ are functions on $x_0$, $x_1$, $x_2$, $x_3$.
\[
1)  y=ax, \ z=dx,\vspace{1mm} \
\varphi_{yy}-\varphi_{zz}=F(\varphi);
\vspace{1mm} \\
\]
\[
2) y=ax, \ z=\left((bx)^2+(cx)^2+(dx)^2\right)^{1/2}, \vspace{1mm}
\ \varphi_{yy}-\varphi_{zz}-\frac{2}{z}\varphi_z=F(\varphi);
\vspace{1mm} \\
\]
\[
3) y=bx+\Phi(ax+dx), \ z=cx, \vspace{1mm} \
-\varphi_{zz}-\varphi_{yy}=F(\varphi);
\vspace{1mm} \\
\]
\[
4) y=\left((bx)^2+(cx^2)\right)^{1/2}, \ z=ax+dx, \vspace{1mm} \
  -\varphi_{yy}-\frac{1}{y}\varphi_{y}=F(\varphi).
\]

\end{document}